\documentclass[twocolumn, prl, aps, superscriptaddress, longbibliography,showpacs,amsmath,amssymb,floatfix]{revtex4-1}
\usepackage{changes}
\usepackage{graphicx}
\usepackage{dcolumn}
\usepackage{bm}
\usepackage{graphicx}                                   
\usepackage{amssymb}

\usepackage{amsmath}
\usepackage{epsfig}
\usepackage{xcolor}
\usepackage{tabu}
\usepackage{mathtools}
\usepackage[colorlinks,linkcolor=blue,anchorcolor=blue,citecolor=blue,urlcolor=blue]{hyperref}
\usepackage{physics}
\usepackage{float}
\usepackage{diagbox}
\usepackage{inputenc}

\begin{document}
    
\title{Additive-Functional Approach to Transport in Periodic and Tilted Periodic Potentials}
\author{Sang Yang}
\email{yangsang@mail.ustc.edu.cn}
\affiliation{University of Science and Technology of China, Hefei, 230026, China}
\author{Zhixin Peng}
\affiliation{University of Science and Technology of China, Hefei, 230026, China}


\date{\today}
	
\begin{abstract}

In this Letter, we clarify the physical origin of effective transport in periodic and tilted periodic systems. When Brownian dynamics is examined on the scale of a single period, the particle displacement admits a natural separation into a bounded part associated with recurrent motion within the periodic landscape, and an unbounded stochastic part that grows in time and carries the net transport. We show that effective drift and diffusion are governed entirely by this unbounded component, while local potential-induced fluctuations contribute only bounded corrections. Treating the displacement as an additive functional of the stochastic dynamics provides a rigorous formulation of this separation and leads to a corrector–martingale representation at the trajectory level. Within this framework, classical results—including the Lifson–Jackson formula for unbiased periodic systems and the Stratonovich expressions for tilted periodic potentials—follow as direct consequences of the same underlying structure. The same perspective extends naturally to higher-dimensional periodic environments, recovering the standard homogenized transport tensors.

\end{abstract}
\maketitle

Transport in periodic energy landscapes is a central theme in nonequilibrium statistical physics, with applications ranging from diffusion in solids \cite{frenken1985observation} and soft matter \cite{ma2015colloidal,castaneda2025colloidal} to Josephson junctions \cite{barone1982physics}, driven charge-density waves \cite{gruner1981nonlinear}, superionic conductors \cite{fulde1975problem}, rotating dipoles in external fields \cite{reguera2000controlling}, and synchronization phenomena \cite{lindsey1972synchronization}. Since the pioneering work of Lifson and Jackson \cite{lifson1962self}, effective transport in periodic systems has been recognized as a genuinely macroscopic phenomenon. On sufficiently large scales, the motion becomes diffusive again, but with renormalized drift and diffusion coefficients that encode the influence of the microscopic potential landscape. Classical derivations of these transport coefficients have therefore focused on macroscopic observables, most notably stationary probability distributions \cite{gunther1979mobility}, mean first-passage times across a unit cell \cite{lifson1962self,weaver1979effective}, or solutions of the Fokker–Planck equation under appropriate boundary conditions \cite{defaveri2023brownian}. While these approaches have led to a variety of elegant and powerful results—most prominently the Lifson–Jackson formula for unbiased periodic potentials and the Stratonovich expressions \cite{reimann2001giant,reimann2002brownian,reimann2002diffusion} for tilted systems—they treat different cases within distinct technical frameworks. As a consequence, the physical origin of effective transport itself is often taken for granted rather than explicitly addressed. Namely, what microscopic feature of the stochastic dynamics is responsible for the emergence of well-defined macroscopic drift and diffusion, and whether this mechanism admits a unified description across different transport scenarios.

A common feature of periodic transport problems is that the microscopic dynamics is recurrent on the scale of a single period, whereas the particle displacement grows without bound on macroscopic scales. Standard approaches capture this separation only indirectly through coarse-grained observables. A more direct viewpoint is to treat the particle displacement itself as the primary object of interest. From this perspective, the key difficulty is to disentangle the bounded fluctuations arising from repeated exploration of the periodic landscape from the unbounded component responsible for long-range transport. Without such a separation, it remains unclear how local potential-induced motion contributes to macroscopic drift and diffusion, or why different periodic transport scenarios nevertheless admit well-defined effective coefficients.

\begin{figure}
    \centering
    \includegraphics[width=0.8\linewidth]{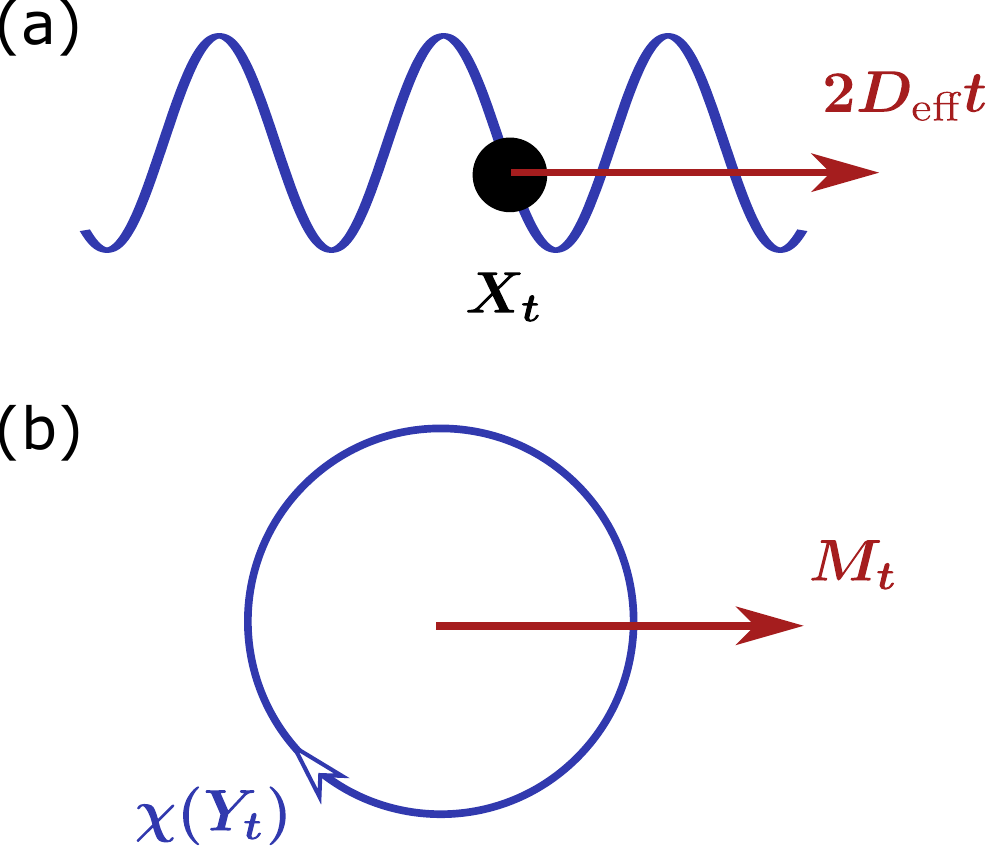}
    \caption{Effective transport in periodic systems and its trajectory-level origin. (a) Brownian motion in a periodic potential $U(x)$. $X_t$ is the displacement. Mean-square displacement grows linearly in time on large scales, which defines an effective diffusion coefficient $D_{\text {eff }}$. (b) Trajectory-level decomposition of the particle displacement. Folding the dynamics onto a single unit cell defines a recurrent process $Y_t=X_t \bmod L$, which repeatedly explores the same periodic landscape. The displacement $X_t$ is separated into a bounded contribution associated with intra-cell recurrent motion $\chi(Y_t)$ and an unbounded stochastic contribution $M_t$.}
    \label{fig-fig1}
\end{figure}

In this Letter, we show that this separation emerges naturally when the displacement is treated as an additive functional of the stochastic dynamics. At the trajectory level, the displacement can be decomposed into a bounded contribution associated with recurrent motion within a single period and an unbounded stochastic contribution that accumulates over successive periods (see in Fig. \ref{fig-fig1}(b)). This leads to a corrector–martingale representation, in which the bounded part encodes local, potential-induced fluctuations, while the martingale part fully determines macroscopic transport. From this viewpoint, effective drift and diffusion follow directly from the long-time statistics of the unbounded component of the displacement, rather than from properties of the stationary distribution alone.

Within this framework, classical results for periodic transport acquire a unified interpretation. In unbiased periodic systems, the absence of a stationary current implies that the unbounded component of the displacement is purely diffusive, yielding the Lifson–Jackson formula. When a constant bias is applied, the same decomposition remains valid, but the unbounded contribution carries a nonzero mean, reflecting the presence of a finite probability current; both the effective drift and diffusion then follow from the same structure and recover the Stratonovich expressions for tilted periodic potentials. Periodic and tilted transport thus correspond to zero-current and finite-current realizations of a single trajectory-level mechanism. Because the construction is formulated on a periodic unit cell, it extends naturally to higher-dimensional periodic environments, leading to the standard homogenized transport tensors.

Finally, we emphasize that the trajectory-level separation underlying our approach is not restricted to constant diffusivity or one-dimensional models. When the diffusivity varies periodically in space, the same additive-functional structure persists, with modified stationary measures and cell problems accounting for spatially heterogeneous mobility. In all cases, the bounded–unbounded decomposition identifies the unbounded stochastic component as the sole carrier of macroscopic transport. By focusing on the structure of particle displacements rather than on specific solution techniques, the present framework provides a physically transparent and unified starting point for analyzing effective transport in periodic media.

{\it Additive-functional decomposition of particle displacement.} We start from overdamped Brownian motion in a periodic landscape $U(x+L)=U(x)$ with constant diffusivity $D_0$,
\begin{equation}
    d X_t=b\left(X_t\right) d t+\sqrt{2 D_0} d W_t, \  b(x)=\mu\left(F-U^{\prime}(x)\right),
    \label{eq-eq1}
\end{equation}
and fold the dynamics onto a single unit cell by $Y_t=X_t \bmod L \in[0, L)$. The folded process is ergodic on the torus and admits a stationary density $\rho$.

Integrating the SDE gives the displacement in the standard additive-functional form
\begin{equation}
    X_t-X_0=\int_0^t b\left(Y_s\right) d s+\sqrt{2 D_0} W_t.
    \label{eq-eq2}
\end{equation}
The first term is an additive functional of the stationary cell dynamics, while the second is a martingale. Effective transport is encoded in the long-time behavior of this additive functional, but in its present form the drift and diffusion are entangled with bounded intra-cell fluctuations.

To extract the transport part, we introduce a periodic corrector $\chi(x)$ on the unit cell and rewrite the displacement as a sum of a bounded term plus a purely stochastic accumulation. The corrector is defined as the solution of the cell (Poisson) problem
\begin{equation}
    \mathcal{L} \chi(x)=v-b(x), \ \langle\chi\rangle_\rho=0, \ \mathcal{L}=b(x) \partial_x+D_0 \partial_x^2,
    \label{eq-eq3}
\end{equation}
where the solvability condition fixes the effective drift
\begin{equation}
    v=\langle b\rangle_\rho.
    \label{eq-eq4}
\end{equation}
This choice ensures that the right-hand side of Eq. (\ref{eq-eq3}) has zero $\rho$-mean, so a periodic solution exists and is unique up to an additive constant.

Applying Itô's lemma to $\chi\left(Y_t\right)$ yields
\begin{equation}
    d \chi\left(Y_t\right)=\mathcal{L} \chi\left(Y_t\right) d t+\sqrt{2 D_0} \chi^{\prime}\left(Y_t\right) d W_t,
    \label{eq-eq5}
\end{equation}
and inserting Eq. (\ref{eq-eq3}) gives
\begin{equation}
    \chi\left(Y_t\right)-\chi\left(Y_0\right)=\int_0^t\left(v-b\left(Y_s\right)\right) d s+\sqrt{2 D_0} \int_0^t \chi^{\prime}\left(Y_s\right) d W_s,
    \label{eq-eq6}
\end{equation}

Subtracting Eq. (\ref{eq-eq6}) from Eq. (\ref{eq-eq2}) yields the trajectory-level separation
\begin{equation}
    X_t-X_0-v t=\underbrace{\chi\left(Y_0\right)-\chi\left(Y_t\right)}_{\text {bounded }}+\underbrace{M_t}_{\text {unbounded }},
    \label{eq-eq7}
\end{equation}
with the martingale
\begin{equation}
    M_t=\sqrt{2 D_0} \int_0^t\left[1+\chi^{\prime}\left(Y_s\right)\right] d W_s.
    \label{eq-eq8}
\end{equation}
Because $\chi$ is periodic, $\chi\left(Y_0\right)-\chi\left(Y_t\right)$ remains bounded for all $t$, so long-time transport is governed entirely by the unbounded martingale term.

The quadratic variation of $M_t$ is
\begin{equation}
    \langle M\rangle_t=2 D_0 \int_0^t\left[1+\chi^{\prime}\left(Y_s\right)\right]^2 d s,
    \label{eq-eq9}
\end{equation}
and ergodicity of $Y_t$ gives the effective diffusion coefficient
\begin{equation}
\begin{aligned}
    D_{\text {eff }}&=\lim _{t \rightarrow \infty} \frac{\operatorname{Var}\left(X_t-v t\right)}{2 t}\\
    &=\frac{1}{2} \lim _{t \rightarrow \infty} \frac{\mathbb{E}\langle M\rangle_t}{t}=D_0\left\langle\left[1+\chi^{\prime}(x)\right]^2\right\rangle_\rho .
    \label{eq-eq10}
\end{aligned}
\end{equation}
Eq. (\ref{eq-eq3}-\ref{eq-eq10}) provide a reusable recipe. Once the stationary cell dynamics is identified, one solves a unit-cell Poisson problem for $\chi$, and then reads off $v$ and $D_{\text {eff }}$ from the martingale part of the displacement.

{\it Unbiased periodic systems: zero current and effective diffusion.} We first apply the above workflow to unbiased periodic systems with $F=0$. In this case the dynamics satisfies detailed balance and admits a reversible stationary density on the unit cell,
\begin{equation}
    \rho_0(x)=\frac{e^{-\beta U(x)}}{\int_0^L e^{-\beta U(y)} d y}, \  J=b(x) \rho_0(x)-D_0 \rho_0^{\prime}(x)=0 .
    \label{eq-eq11}
\end{equation}
The absence of a stationary current has two immediate consequences within the additive-functional framework. First, the effective drift vanishes,
\begin{equation}
    v=\langle b\rangle_{\rho_0}=0,
    \label{eq-eq12}
\end{equation}
so macroscopic transport reduces to pure diffusion. Second, the cell dynamics is purely recurrent: the particle repeatedly explores the same potential landscape without systematic bias, and all directed motion arises solely from stochastic fluctuations accumulated over many periods.

This means that the martingale $M_t$ in Eq. (\ref{eq-eq7}) has zero mean, while its quadratic variation determines the long-time spreading of trajectories. Solving the corresponding unit-cell Poisson problem,
\begin{equation}
    \mathcal{L} \chi(x)=-b(x), \quad\langle\chi\rangle_{\rho_0}=0,
    \label{eq-eq13}
\end{equation}
and substituting the solution into Eq. (\ref{eq-eq10}) yields the effective diffusion coefficient in the LifsonJackson form,
\begin{equation}
    D_{\mathrm{eff}}=\frac{D_0}{\left\langle e^{\beta U}\right\rangle_L\left\langle e^{-\beta U}\right\rangle_L}.
    \label{eq-eq14}
\end{equation}
This expression admits a clear physical interpretation. The factor $\left\langle e^{\beta U}\right\rangle$ reflects the suppression of long-range transport by energy barriers, while $\left\langle e^{-\beta U}\right\rangle$ accounts for enhanced residence times in potential wells. Importantly, both contributions enter symmetrically because local potential-induced motion affects transport only through bounded corrections absorbed by the corrector $\chi$; the macroscopic diffusion is entirely governed by the unbounded stochastic accumulation encoded in the martingale. From this viewpoint, the Lifson-Jackson formula is not a special property of equilibrium distributions, but a direct consequence of the trajectory-level separation inherent to periodic dynamics.

{\it Tilted periodic systems: finite current and biased transport.} We now turn to tilted periodic systems with a constant external force $F \neq 0$. In contrast to the unbiased case, detailed balance is broken and the stationary state on a unit cell carries a finite probability current $J$. Writing the tilted potential as
\begin{equation}
    V(x)=U(x)-F x,
    \label{eq-eq15}
\end{equation}
the stationary density satisfies
\begin{equation}
    J=b(x) \rho_F(x)-D_0 \rho_F^{\prime}(x)=\text { const },
    \label{eq-eq16}
\end{equation}
and can be expressed in the standard flux-carrying form
\begin{equation}
    \rho_F(x)=\frac{1}{\mathcal{N}} e^{-\beta V(x)} \int_x^{x+L} e^{\beta V(y)} d y,
    \label{eq-eq17}
\end{equation}
with normalization $\mathcal{N}=\int_0^L d x I_{+}(x)$ and
\begin{equation}
    I_{+}(x)=e^{-\beta V(x)} \int_x^{x+L} e^{\beta V(y)} d y.
    \label{eq-eq18}
\end{equation}
Within the additive-functional framework, the presence of a finite current has a direct and transparent consequence. The solvability condition of the Poisson problem now yields a nonzero effective drift,
\begin{equation}
    v=\langle b\rangle_{\rho_F}=J L,
    \label{eq-eq19}
\end{equation}
reflecting the systematic bias accumulated as the particle repeatedly crosses successive periods. The displacement decomposition (7) remains valid, but the martingale $M_t$ now carries a nonzero mean in addition to its fluctuating part.

Solving the corresponding unit-cell Poisson problem for the corrector $\chi$ and inserting the result into the general expression (10) yields the effective diffusion coefficient. Introducing the auxiliary function
\begin{equation}
    I_{-}(x)=e^{\beta V(x)} \int_{x-L}^x e^{-\beta V(z)} d z,
    \label{eq-eq20}
\end{equation}
the result can be written in the compact Stratonovich form
\begin{equation}
    D_{\mathrm{eff}}=D_0 L^2 \frac{\int_0^L d x I_{+}(x)^2 I_{-}(x)}{\left(\int_0^L d x I_{+}(x)\right)^3}.
    \label{eq-eq21}
\end{equation}
The physical content of this expression is most transparent when compared with the unbiased case. Local motion within a single period still produces only bounded corrections absorbed by the corrector $\chi$, while both the effective drift and diffusion originate entirely from the unbounded accumulation encoded in the martingale. The qualitative difference is that, in the presence of a finite current, the martingale acquires a nonzero mean, leading to biased transport. In this sense, periodic and tilted periodic systems correspond respectively to zero-current and finite-current realizations of the same trajectory-level separation, rather than to distinct transport mechanisms.

{\it Higher-dimensional periodic environments.} The additive-functional structure identified above extends directly to diffusion in higherdimensional periodic environments. Treating each component of the displacement as an additive functional of the cell dynamics leads to a vector-valued corrector and a martingale decomposition analogous to Eq. (\ref{eq-eq7}). The effective drift is given by the stationary average $\mathbf{v}=\langle\mathbf{b}\rangle_\rho$, while the effective diffusion tensor follows from the martingale covariance and takes the standard homogenized form \cite{pavliotis2008diffusive,duncan2023brownian}
\begin{equation}
    \left(D_{\mathrm{eff}}\right)_{i j}=D_0\left\langle\left(\mathbf{e}_i+\nabla \chi_i\right) \cdot\left(\mathbf{e}_j+\nabla \chi_j\right)\right\rangle_\rho
    \label{eq-eq22}
\end{equation}
Thus, higher-dimensional periodic transport does not require new mechanisms: the corrector absorbs bounded intra-cell motion, and macroscopic transport is fully controlled by the unbounded stochastic accumulation. Details are given in the Supplemental Material.

{\it Periodically varying diffusivity.} When the diffusivity varies periodically in space, $D(x+L)=D(x)$, the displacement remains an additive functional of the stochastic dynamics, and the same corrector-martingale structure applies. The effective drift is still determined by the stationary average $v=\langle b\rangle_\rho$, while the effective diffusion coefficient becomes
\begin{equation}
    D_{\text {eff }}=\left\langle D(x)\left[1+\chi^{\prime}(x)\right]^2\right\rangle_\rho
    \label{eq-eq23}
\end{equation}
Spatial variations of $D(x)$ modify the weighting of fluctuations inside the unit cell but do not alter the origin of macroscopic transport, which remains governed by the unbounded stochastic accumulation. Explicit expressions are provided in the Supplemental Material.

In summary, we have developed a unified additive-functional framework for transport in periodic and tilted periodic potentials. By decomposing the particle displacement into a bounded periodic corrector and a martingale component, we identified the latter as the sole contributor to long-time drift and diffusion. This approach naturally recovers the Lifson–Jackson diffusion formula in reversible periodic systems and yields the Stratonovich expressions for the effective drift and diffusion in biased periodic potentials, where detailed balance is broken and a finite stationary current emerges. Beyond providing transparent derivations of classical results, the present formulation clarifies the common mathematical structure underlying equilibrium and nonequilibrium transport in periodic media. The additive-functional perspective offers a systematic and extensible route to transport coefficients in more general settings, including higher-dimensional periodic structures, spatially varying mobilities, and nonequilibrium environments, and thus establishes a versatile bridge between stochastic-process theory and nonequilibrium statistical physics.

\bibliographystyle{IEEEtran}
\bibliography{ref.bib}
\end{document}